\documentclass[twocolumn,floats,prl,showpacs,aps]{revtex4}
\usepackage[dvips]{graphicx}
\usepackage{textcomp}
\usepackage[T1]{fontenc}

\begin{document}

\title{Disorder effects in pnictides\,: a tunneling spectroscopy study}
\author{Y. Noat}
\email{yves.noat@insp.jussieu.fr}
\author{T. Cren}
\author{V. Dubost}
\author{S. Lange}
\author{F. Debontridder}
\author{D. Roditchev}
\affiliation{Institut des Nanosciences de Paris, CNRS UMR 7588,
Universit\'e Pierre et Marie Curie Paris 6, Campus Boucicaut, 140
rue de Lourmel, F-75015 Paris, France}

\author{J. Marcus}
\author{P. Toulemonde}
\affiliation{Institut N\'eel, CNRS et Universit\'e Joseph Fourier,
25 rue des Martyrs, BP 166, F-38042 Grenoble, France}

\author{W. Sacks}
\affiliation{Institut de Min\'eralogie et de Physique des Milieux Condens\'es,
CNRS UMR 7590, Campus Boucicaut, 140 rue de Lourmel, F-75015 Paris, France}

\date{\today}

\begin{abstract}

We present the synthesis and the tunneling spectroscopy study of
superconducting FeSe$_{0.5}$Te$_{0.5}$ ($T_c=14$\,K),
SmFeAsO$_{0.85}$ ($T_c=45$\,K) and SmFeAsO$_{0.9}$F$_{0.1}$
($T_c=52$\,K). The samples were characterized by Rietveld refinement
of X-ray diffraction patterns and transport measurements. Tunneling
experiments on FeSe$_{0.5}$Te$_{0.5}$ revealed a single
superconducting gap $\sim1$\,meV in BCS-like tunneling conductance
spectra. In SmFeAsO$_{0.85}$ and SmFeAsO$_{0.9}F_{0.1}$ however,
more complex spectra were observed characterized by two gap-like
structures at $\sim4$\,meV and $\sim10\,$meV. These spectra are
qualitatively understood assuming a two-band superconductor with a
'\textbf{s$\pm$}' order parameter. We show that depending on the
sign relation between the pairing amplitudes in the two bands, the
interband quasiparticle scattering has a crucial effect on the shape
of the tunneling spectra. Detailed analysis of the tunneling
spectroscopy data supports the '\textbf{s$\pm$}' model, but
single-gap spectra found in FeSe$_{0.5}$Te$_{0.5}$ are more
compatible with a disorder-induced 's'-wave gap, due to the Se-Te
substitution.
\end{abstract}

\pacs{74.25.Gz, 74.72.Jt, 75.30.Fv, 75.40.-s}

\maketitle

\section{Introduction}

In 2008, i.e. 22 years after Bednorz and Muller discovery
\cite{Bednorz} of superconductivity in cuprates, the exciting fining
of superconductivity in LaFeAsO$_{1-x}$F$_x$ ($T_c=23$\,K, reaching
$T_c=43$\,K under pressure) \cite{Kamihara,Takahashi}, gave rise to
a completely new family of high-T$_c$ superconductors, with the
critical temperature culminating at $T_c=55$\,K in
SmFeAsO$_{0.9}F_{0.1}$ \cite{Sm-1111}. Understanding these materials
is of fundamental importance, as one could expect a 'new' and
unconventional type of superconductivity to occur there. There
exists also a hope in the scientific community that understanding
the nature of pairing in iron-based materials could shed light on
the complex superconducting (SC) state of the high-T$_c$ cuprates
(see the review \cite{review_cuprates} and refs. therein) where the
mechanism is still not fully understood.

As in the cuprates, the iron pnictides have a layered structure and
are close to a magnetic transition. The parent compound exhibits
magnetism which is destroyed by electron/hole doping, and the SC
state ensues. Another analogy with the cuprates is the possible role
of strong electronic correlations. However, the parent compound for
oxypnictides seems to be a correlated metal rather than an
antiferromagnetic Mott insulator. Superconductivity is obtained in
this system by charge doping either outside of the FeAs blocks (by O
deficiency or F substitution on the O site in the LaFeAsO compound)
or directly in the active FeAs planes (by Co substitution for Fe in
BaFe$_2$As$_2$, see for example Sefat et al. \cite{Sefat_Co_PRL}).
In addition, superconductivity can be induced by mechanical pressure
in such systems
\cite{Torikachvili,Okada,Alireza,Matsubayashi,Igawa}, i.e. without
introducing chemical disorder.

There is thus a number of questions to be answered. Is the SC state
in these compounds unconventional? If yes, what is the SC gap
symmetry and the underlying pairing mechanism? Are the
iron-pnictides an example of superconductivity mediated by
magnetism? What is the nature of the normal state? Does a pseudogap
exist in the normal state, i.e. just above $T_c$, and in the vortex
core?

In this paper, we describe the fabrication of SmFeAsO$_{0.85}$
($T_c^{mid}$=45\,K, \cite{MID}), SmFeAsO$_{0.9}F_{0.1}$
($T_c^{mid}$=52\,K) oxypnictide crystals and FeSe-based SC
FeSe$_{0.5}$Te$_{0.5}$ ($T_c^{mid}$=12\,K) materials. The samples
were characterized by temperature-dependent resistivity
measurements, Rietveld refinement of X-ray diffraction patterns and
finally studied by tunneling spectroscopy (TS) at 4.2~K. While the
signature of a single 'BCS-like' gap was observed in tunneling
spectra of FeSe$_{0.5}$Te$_{0.5}$, two 'gap-like' structures, at
$\sim 3$\,meV and $\sim 10$~meV, were revealed in the tunneling
conductance spectra of SmFeAsO$_{0.85}$ and SmFeAsO$_{0.9}F_{0.1}$.
These structures are interpreted as two SC gaps. Moreover, some
local tunneling conductance spectra showed a well-pronounced
zero-bias peak - a structure which is absent in the tunneling DOS of
conventional superconductors, but observed in 'd'-wave high-T$_c$
cuprates. In order to give a qualitative understanding of these
results, we numerically studied the two-band superconductor in the
framework of Schopohl and Scharnberg model \cite{Schopohl}. We
focused on the influence of interband quasiparticle (QP) scattering
on the tunneling DOS in two distinct cases\,: \textit{i} - pairing
via magnetic coupling ('\textbf{s$\pm$}' model \cite{Mazin}),
resulting in a sign reversal between two parts of the SC order
parameter (OP) corresponding to two different bands; \textit{ii} -
phonon pairing, with two gaps in the OP having the same phase.
Remarkably, we found the interband QP scattering to affect very
differently the SC DOS, qualitatively and quantitatively,  in these
two cases. Within '\textbf{s$\pm$}' model, the QP peaks in the DOS
are strongly washed out by the interband QP scattering, while within
's'-wave model, they are much less affected. This difference allowed
us to discriminate between two models and to support the
'\textbf{s$\pm$}' scenario in the case of SmFeAsO$_{0.85}$ and
SmFeAsO$_{0.9}F_{0.1}$.

\section{I. Pnictides and iron chalcogenides: Structure, Doping and Superconductivity}

When the SC state is obtained by charge doping of the active FeAs
planes, the rare-earth element -- oxygen (RE-O) plane acts as a
charge reservoir, controlling the carrier concentration in the FeAs
planes via the substitution of oxygen by fluorine or via oxygen
vacancies. Remarkably, spin fluctuations were observed in
SmFeAsO$_{1-x}F{_x}$ (with x=0.18 and 0.3) near the SC
transition\,\cite{Drew}, and magnetic resonance was reported in
FeSe$_{x}$Te$_{1-x}$ \cite{MagnRes_FeSe}. The superconducting OP was
shown to be coupled to a magnetic resonance in
Ba$_{0.6}$K$_{0.4}$Fe$_2$As$_2$ \cite{Christianson}. These
experiments suggest that, in possible analogy to the cuprates,
magnetism plays a crucial role in the formation of the SC state.
Furthermore, it has been argued that the attractive interaction is
mediated by spin fluctuations \cite{Mazin} instead of phonons. As
explored in this work, this could lead to a change of sign of the OP
between different Fermi sheets\,\cite{Mazin}.

Arsenic high toxicity has motivated the search for other materials
with similar structural properties. This led to the discovery of
superconductivity in the relatively simple compound FeSe \cite{Hsu}
with a critical temperature $T_c=8$\,K, but posing the additional
question of chemical disorder. With their similar layered
structures, the comparison of the FeAs and FeSe families should be
fruitful. As mentioned above, SmFeAsO$_{1-x}$ can be doped either by
O deficiency or by F substitution and leaving the FeAs planes
pristine. To the contrary, in the FeSe system, where the SC is
induced by Se deficiency or by substitution of Se by Te atoms, two
different effects are obtained\,: In the first case, the additional
Fe atoms are intercalated between the FeSe planes which thus remain
intact upon substitution. The critical temperature reaches 37\,K
\cite{Medvedev_FeSe37K,Margadonna_FeSe37K,Sidorov} or a slightly
lower value \cite{Garbarino,Mizuguchi,Miyoshi} under applied
pressure. In the second case, when doping by substitution, the Te
atoms are introduced within the FeSe planes and the critical
temperature is much lower\,: $\sim$14\,K.

To address, at least partially, the question of the SC OP in iron
pnictides, we performed tunneling spectroscopy (TS) in the Scanning
Tunneling Microscope (STM) geometry. This powerful technique allows
local measurements of the QP excitation spectra of superconductors.
Indeed, if one of the electrodes is a normal metal (with a
constant density of state $N_N(E)\approx \textit{const}$) and the second
one is a superconducting material characterized by an excitation
spectrum $N_S(E)$, the derivative of the tunneling current $I(V)$
across such a NIS junction as a function of bias voltage $V$ reads:
\begin{equation}
dI(V)/dV\ \propto\ \int\ N_{S}\,(E+eV) g(E) dE
\label{eq_didv}
\end{equation}
where $g(E)$ is the derivative of the Fermi-Dirac function $f(E)$:
$g(E)=-\partial f(E)/\partial E$; $g(E)$ is a bell-shaped function
of width $\sim 3.5\,kT$. The spectroscopic resolution of the TS is
therefore mainly limited by the temperature of the experiment. Since
the pioneering experiments by Giaever \cite{Giaever} who measured
the SC gap in Pb in planar tunneling junctions, the TS in STM
geometry has been successfully used to study conventional (see, for
instance, the extensive contributions of Hess et al. \cite{Hess}) as
well as high-Tc superconductors (for a review of STS in cuprates,
see \cite{Review_Fisher} and refs. therein).

The above commonly used expression for the tunneling conductance
$dI(V)/dV$ neglects the $\textbf{k}$-dependence of the tunneling
transparency. Though, the tunneling towards the materials with
complex band structure may result in different tunneling
probabilities for different parts/sheets of the Fermi surface. The
measured tunneling conductance spectra thus probe different portions
of the Fermi surface with different spectral weights. This effect,
first observed in cuprates \cite{Mallet}, was crucial in
understanding the two-band superconductivity in MgB$_2$
\cite{Giubileo}, but also in CaC$_6$ \cite{Bergeal} and recently, in
Ba$_8$Si$_{46}$ \cite{Noat_Ba8Si46}. Thus, the tunneling density of
states (DOS) may significatively differ from the QP excitation
spectrum of the bulk material.

Despite the large number of experimental studies of FeAs and
FeSe-based superconductors, there are relatively few comprehensive
TS results. The case of Fe(Se,Te) is maybe the clearer. Indeed, two
scanning tunneling spectroscopy studies showed a 's'-wave like
single gap superconducting density of states in this material
\cite{Kato_FeSeTe_2009, New Science}. In their scanning tunneling
spectroscopy study of cleaved surfaces of single crystals of the
iron chalcogenides Fe$_{1.05}$Se$_{015}$Te$_{0.85}$, Kato et al.
\cite{Kato_FeSeTe_2009} found a simple BCS 's'-wave OP, relatively
homogeneous, on a spatially inhomogeneous spectral background.
Similarly, Hanaguri et al. \cite{New Science} found a clear s-wave
like density of state in Fe(Se,Te). Nevertheless, basing on
quasi-particle scattering interference patterns studies, the authors
claim to have evidenced '\textbf{s$\pm$}' superconductivity in
Fe(Se,Te).

The situation is more complex for FeAs-based materials where more
complexe signatures where observed. Millo et al. \cite{Millo}
studied polycrystalline samples of SmFeAsO$_{0.85}$ and, in some
regions, observed a gap $\sim$ 8\,meV with clear QP peaks; the shape
of the tunneling spectra was found compatible with 'd' wave symmetry
of the SC OP. In other regions of the sample, a zero bias peak in
the conductance spectra was observed. Although the authors suggested
a possible 'd'-wave OP, they also remarked that, on a large part of
the surface ($\sim 70 \%$), TS spectra with a characteristic 'V'
shaped background in tunneling conductance were dominant (the
conductance increasing almost linearly with the bias voltage,
$dI(V)/dV \propto \left| V \right|$). Spectra of a very
similar shape have already been observed in YBa$_2$Cu$_3$O$_7$
\cite{Cren_EPL2000} and other cuprates, including the electron-doped
case Sm$_{1.85}$Ce$_{0.15}$CuO$_4$ \cite{Zimmers}.

Pan et al. \cite{Pan} studied NdFeAs$_{0.86}$F$_{0.14}$ and found
two characteristic gaps in spatially separated locations of the
sample: a small gap of $\sim $9\,meV with a BCS-like shape and
temperature dependence, and a larger one $\sim $18\,meV having a
non-BCS shape. Both gaps were found to close at $T_c$ of the bulk
material. In addition, Pan et al. identified a `pseudogap' opening
just above $T_c$. Yin et al. \cite{Yin} have observed a gap with
well pronounced QP peaks in all locations of studied
BaFe$_{1.8}$Co$_{0.2}$As$_2$ sample. The SC nature of the observed
gap was proven by vortex imaging under magnetic field; a disordered
vortex lattice, attributed to bulk pinning, was reported. Chen et
al.\,\cite{Chen} found a single BCS-like gap in
SmFeAsO$_{0.85}$F$_{0.15}$.

Multigap superconductivity, such as reported by Pan et al.
\cite{Pan}, has been suggested for FeAs-based superconductors, as a
possible explanation of reported experimental results. Wang et al.
\cite{Wang_PC} explained in this way the point contact spectroscopy
data obtained in SmFeAsO$_{1-x}$F$_{x}$ for $x=0.1$ and by Daghero
et al. \cite{Daghero} for $x=0.2$ and $x=0.09$. Indeed, Wang et al.
\cite{Wang_PC} found two gaps, a small one $\Delta_1=3.7$\,meV and a
large gap $\Delta_2=10.5$\,meV while Daghero et al. \cite{Daghero}
found $\Delta_1=6.15$\,meV and $\Delta_2=18$\,meV for $x=0.2$, and
$\Delta_1=4.9$\,meV and $\Delta_2=15$\,meV for $x=0.09$ (the typical
gap uncertainties being $\sim 10 \%$).

Most of these experiments on SmFeAsO compounds suggest a two bands
superconductivity scenario with an important QP interband coupling
\cite{Schopohl}, such as observed in MgB$_2$
\cite{Liu_MgB2,Schmidt}. Nevertheless, taking into account the
presence of magnetic fluctuations in SC iron pnictides,  one has to
consider a more complex situation, where the SC pairing originates
from a magnetic coupling and gives rise to a change of sign between
the two 's'-wave parts of the SC OP, as considered in \cite{Mazin}.

\section{II. Comparative Crystal Structure}

\begin{figure}
\begin{center}
\includegraphics[width=8cm]{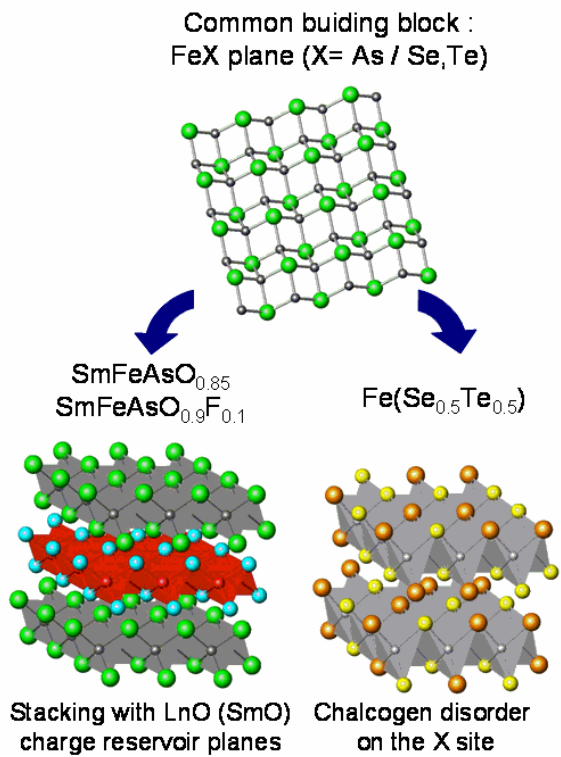}
\end{center}\caption{(Color online)
Building-block approach to the crystal chemistry of iron-based
superconductors. (a) Superconducting FeX (X\,=\,As,\,Se) plane :
Iron(II) is in flattened tetrahedral coordination. (b) Crystal
structure of SmFeAs0$_{0.85}$ and SmFeAs0$_{0.90}$F$_{0.1}$. FeAs
planes are alternatively stacked with LnO charge reservoir planes.
(c) Crystal structure of Fe(Se$_{0.5}$Te$_{0.5}$), showing the
chemical disorder on the X site (X=\,As\,/\,Se,\,Te). For this composition, the
inter-plane excess iron concentration is almost zero. Color code for
(b) and (c). Grey : Fe atoms and Fe-centered polyedra, Green : As
atoms, Blue : Sm atoms, Red : oxygen and oxygen-centered polyedra,
Se atoms (yellow) are randomly substituted by Te atoms (orange). }
\label{structure}
\end{figure}

As previously mentioned, the pnictide SC state is associated with
the FeAs plane, structurally encountered in the natural mineral
mackinawite \cite{Strunz} (Fig.\ref{structure}(a)). A vast series of
compounds can be designed by different stacking of this plane with
other structural units \cite{RenMiniRevue} (Fig.
\ref{structure}(b)). In the case of oxypnictides, such as
SmFeAsO$_{0.85}$ and SmFeAs(O$_{0.9}$F$_{0.1}$), the carrier doping
in the FeAs plane comes from the off-stoichiometry or fluorine
substitution in the RE-O plane. In the first case,
RE-FeAsO$_{1-\delta}$, the effective doping is $2\delta$ electrons
per Fe atom, whereas in the last case RE-FeAs(O$_{1-\delta}$F$_{\delta}$), it
corresponds to $\delta$ electrons. In the latter case, one should note
that the defect is located \emph{outside} the SC plane and
intuitively acts as an additional electrostatic potential. Defects
\emph{within} the SC plane can be induced in this family by the
substitution of iron atoms by cobalt atoms \cite{Sefat_Co,Qi,Wang_Co}.

Concerning the Fe$_{1+y}$Se$_{1-x}$Te$_{x}$ family
(Fig.\ref{structure}(c)), despite a simpler chemical composition,
the crystal chemistry of these materials is rather complex, and one
should carefully consider the off-stoichiometry when discussing the
SC properties \cite{McQueen}. In a perfect stoichiometric compound
all Fe atoms, designated Fe(1), are in the tetrahedral sites of the
planes. When Fe is present in excess (y$>$0), it occupies the sites
between the Fe(Se, Te) planes, designated as Fe(2).
Superconductivity was first discovered in the nominal FeSe$_{0.88}$
Se-deficient compound \cite{Hsu}. It was then demonstrated to
correspond to a nearly stoichiometric compound \cite{Williams}, with
a T$_{c}$ of 8\,K, and then raised to 34-37\,K by applying
hydrostatic pressure
\cite{Medvedev_FeSe37K,Garbarino,Margadonna_FeSe37K,Sidorov}. The
substitution of selenium by tellurium was then examined, allowing
enhancing T$_{c}$ to 14\,K at ambient pressure \cite{Fang}. The
general tendency throughout the isovalent substitution of selenium
by tellurium is the further occupation of the Fe(2) sites between
the planes \cite{Sales}. In this system, the T$_{c}$ first increases
with tellurium substitution with a maximum onset value corresponding
to 14\,K for the composition FeSe$_{0.5}$Te$_{0.5}$ \cite{Wu}. The
tellurium compound Fe$_{1.07}$Te is non-superconducting
\cite{Fruchart}.

At this stage, one should note important differences between the
oxypnictides and iron chalcogenides. First, due to the absence of
charge reservoir planes in the latter case, the inter-plane distance
is smaller (the inter-plane distance, measured with respect to iron
atoms is around 8 \AA\, in the case of oxypnictides, versus 5 \AA\,
in the case of iron chalcogenides) and the three-dimensional
character is more pronounced. Then, the Se/Te substitution occurs
\emph{within} the SC plane, and one could expect strong consequences
on the SC properties. The Se/Te disorder may cause an important QP
scattering that, in the case of a magnetic fluctuation mediated
superconductivity, induces a disorder in the exchange integrals
between the iron atoms and, consequently, the scattering by spin
density waves. Moreover, the additional inter-plane atom Fe(2) is
believed to have a magnetic moment, and its influence to the SC
state could be more complex than a simple electrostatic potential
\cite{Liu}. Finally, we have chosen for this study single crystals
of the composition FeSe$_{0.5}$Te$_{0.5}$, which correspond to both
the highest T$_{c}$ at ambient pressure ($T_c^{mid}$ = 12\,K), a
nearly zero occupation of Fe(2) sites and a maximum Se/Te mixing.

\section{III. Fabrication and characterization of the samples}

\begin{figure}
\begin{center}
\includegraphics[width=7cm]{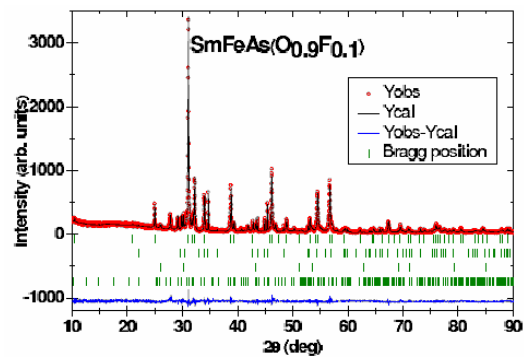}
\end{center}\caption{(Color online)
Rietveld refinement of SmFeAsO$_{0.9}$F$_{0.1}$ sample.
} \label{Rietveld}
\end{figure}

SmFeAsO$_{0.85}$ and SmFeAs(O$_{0.9}$F$_{0.1}$) samples (Sm-1111)
were prepared under high pressure - high temperature using a "belt"
type high pressure cell. Sm, Fe, Fe$_2$O$_3$, As and SmF$_3$ (in the
case of the fluorine doped sample) powders were mixed together and
pressed in the form of cylindrical pellets. They were introduced in
a home-made boron nitride crucible which was surrounded by a
cylindrical graphite resistive heater and the whole assembly was
placed in a pyrophyllite gasket. The samples were treated at 6GPa,
1000-1100$\,^{\circ}{\rm C}$ during 4 hours and then quenched to
room temperature. The XRD patterns show that the major phase is the
one expected, Sm-1111 with small impurities: FeAs, SmAs and
Sm$_2$O$_3$, as illustrated in Fig.\ref{Rietveld}. It shows the
result of the Rietveld refinement for which the structural
parameters of the Sm-1111 phase were refined while only the scaling
factor, lattice parameters and profile shape parameters of the
secondary phases were refined with atomic positions fixed.

\begin{figure}
\begin{center}
\includegraphics[width=5cm]{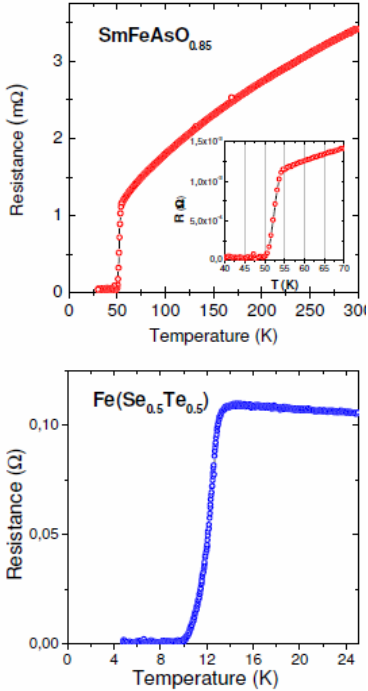}
\end{center}\caption{(Color online)
a) Resistance versus temperature dependence of polycrystalline
SmFeAsO$_{0.85}$ sample (measured by the four terminals method). b)
Resistance versus temperature dependence of a FeSe$_{0.5}$Te$_{0.5}$
single crystal (measured by the four terminals method). }
\label{Resistivity}
\end{figure}

Superconductivity was checked by resistance and a.c. susceptibility
measurements. Both show an onset of superconductivity above 50 K
(Fig.\ref{Resistivity}) but with a larger transition for the
fluorine doped sample ($T_c^{mid}$=45K for SmFeAsO$_{0.9}$F$_{0.1}$)
compared to the compound with oxygen vacancies ($T_c^{mid}$=52K for
SmFeAsO$_{0.85}$).

Single crystals of Fe(Se$_{0.5}$Te$_{0.5}$) were fabricated using
the sealed quartz tube method. The samples were prepared starting
from iron pieces and Se/Te powders mixed together and introduced in
a quartz tube which was then sealed under vacuum. The reactants were
heated slowly at 500$\,^{\circ}{\rm C}$ for 10~h, at
950$\,^{\circ}{\rm C}$ for 5h, slowly cooled down to
350$\,^{\circ}{\rm C}$ at 5$\,^{\circ}{\rm C}$/h and finally
quenched in water. Single crystals were extracted mechanically;
their resistance (Fig.\ref{Resistivity}b) and magnetization versus
temperature were measured. The onset of the SC transition was found
at 14~K ($T_c^{mid}=12$~K) with relatively good transition, i.e. a
zero resistance at 10K.

\section{IV. Tunneling spectroscopy}

\begin{figure}[h]
\begin{center}
\includegraphics[width=5cm]{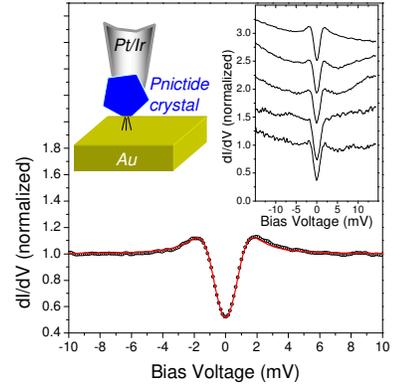}
\end{center}\caption{(Color online)
Circles: Background corrected tunneling conductance spectrum
observed in FeSe$_{0.5}$Te$_{0.5}$ at $T=4.2$~K, tunneling
resistance $R_T=12.5$~M$\Omega$. Solid line: BCS fit using Dynes
formula \cite{Dynes} gives $\Delta=0.95$~meV; $\gamma=0.22$~meV.
Inset (left): Geometry of the tunneling experiment: small SC
crystals were glued on a PtIr tip (and broken in ultra high vacuum
before the experiment in order to get a clean surface). A gold
crystal was used as the counter electrode. Inset (right): tunneling
spectra observed in other studied samples. } \label{FeSeTe}
\end{figure}

We now focus on the TS experiments which were performed in the
inverted junction geometry \cite{Pan,Giubileo}. It is well known
that the surface of iron-based pnictides is very reactive and
contaminates immediately in air. Hence, in order to limit the
contamination the crystals  effects in TS experiments, the samples
were broken in UHV. In practice, small pnictide crystals were glued
to the apex of Pt-Ir STM tips (as illustrated in inset of
Fig.\ref{FeSeTe}) with conducting silver glue and introduced in UHV
chamber containing the STM unit. The crystals were then broken under
ultra-high vacuum conditions. A pure gold surface was used as the
counter-electrode. For each composition a large number of samples
have been studied.

Typical tunneling conductance spectra of FeSe$_{0.5}$Te$_{0.5}$ are
presented in inset of Fig.\ref{FeSeTe}. The spectra clearly show a
single gap at zero bias followed by broad peaks on each side. A weak
spectral background is observed; it varies from one sample to
another. Once this background substracted, the spectra may be well
fitted considering 's'-wave BCS excitation spectrum with finite QP
lifetime (Dynes formula, ref. \cite{Dynes}) as shown on
Fig.\ref{FeSeTe}. The fits of different spectra give a gap value of
$\sim 1.0 \pm 0.1$~meV. Note that it is smaller than the expected value (1.9\,meV) expected for a conventional BCS superconductor for which $2 \Delta /kT_c\approx 3,52$. 
It is important to note the large value of the broadening parameter $\gamma\sim0.2$~meV, varying from one
sample to another. These observations of a single-gap shape of the
spectra are in good agrement with previous tunneling experiments
\cite{Kato_FeSeTe_2009, New Science}. Nevertheless, the gap value
we found is clearly smaller than the 2~mev found by Kato et al.
\cite{Kato_FeSeTe_2009} and the 1,7~meV found by Hanaguri et al.
\cite{New Science}. At this moment, we have no clear explanation for this discrepancy. Thus, while the existence of a single gap in
FeSe$_{0.5}$Te$_{0.5}$ was clearly established, the spectral
broadening did not allow us to distinguish definitively between a
simple 's'-wave or a more complex OP in this materials. Indeed,
basing on quasiparticle interference patterns observed by scanning
tunneling spectroscopy Hanaguri et al. \cite{New Science} suggested
an unconventional '\textbf{s$\pm$}' order parameter for this
material.

\begin{figure}[h]
\begin{center}
\includegraphics[width=8cm]{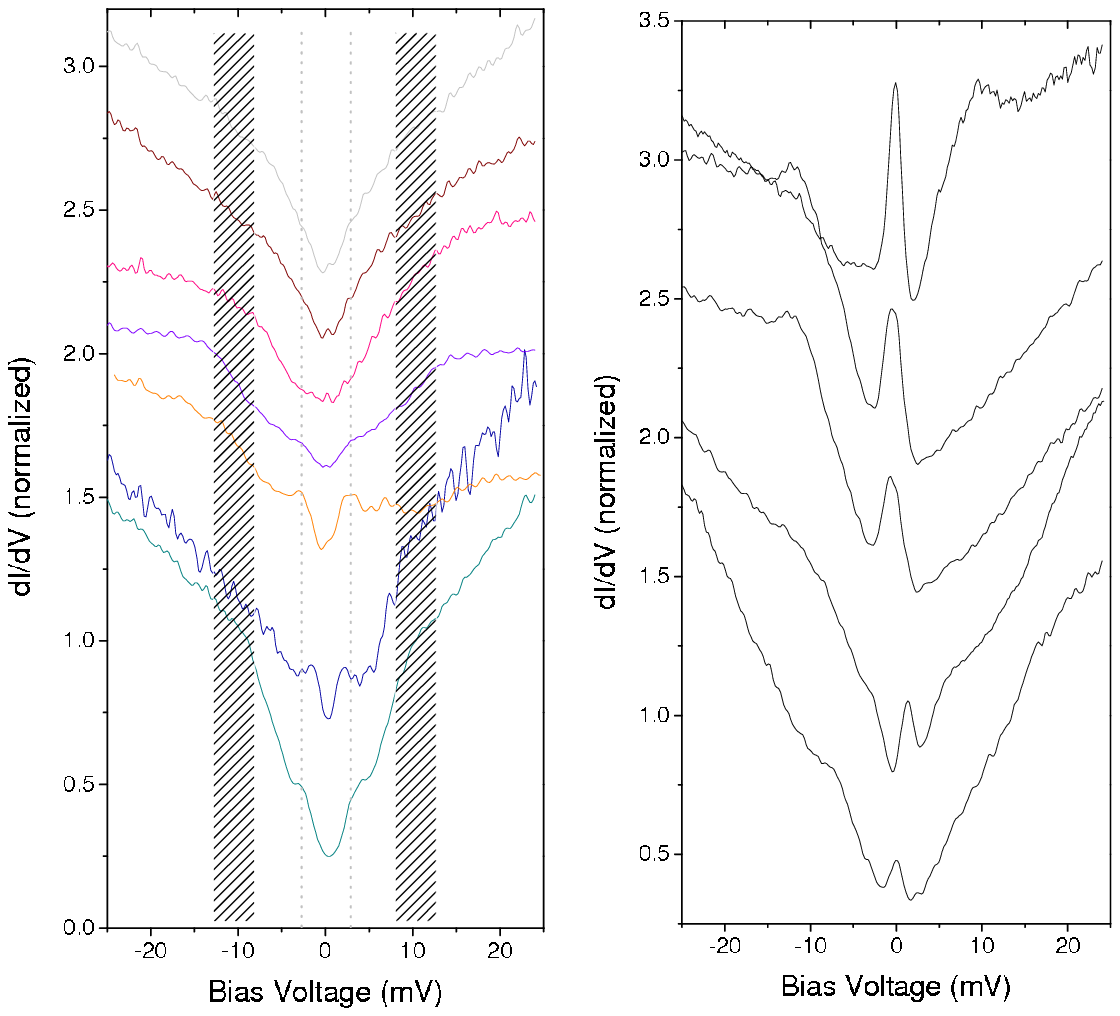}
\end{center}\caption{(Color online)
Tunneling spectra of SmFeAsO$_{0.85}$ acquired at $T=$4.2 and at tunneling resistance $R_T$=12.5M$\Omega$ . Left panel: two gap-like structures at respectively at $\sim $10mV and $\sim $3mV a) superimposed on a 'V' shaped background are clearly visible. Dashed lines and bands are put at characteristic bias values. Right panel: selected spectra exhibiting a zero bias peak.} \label{SmFeAsO}
\end{figure}

\begin{figure}[h]
\begin{center}
\includegraphics[width=8cm]{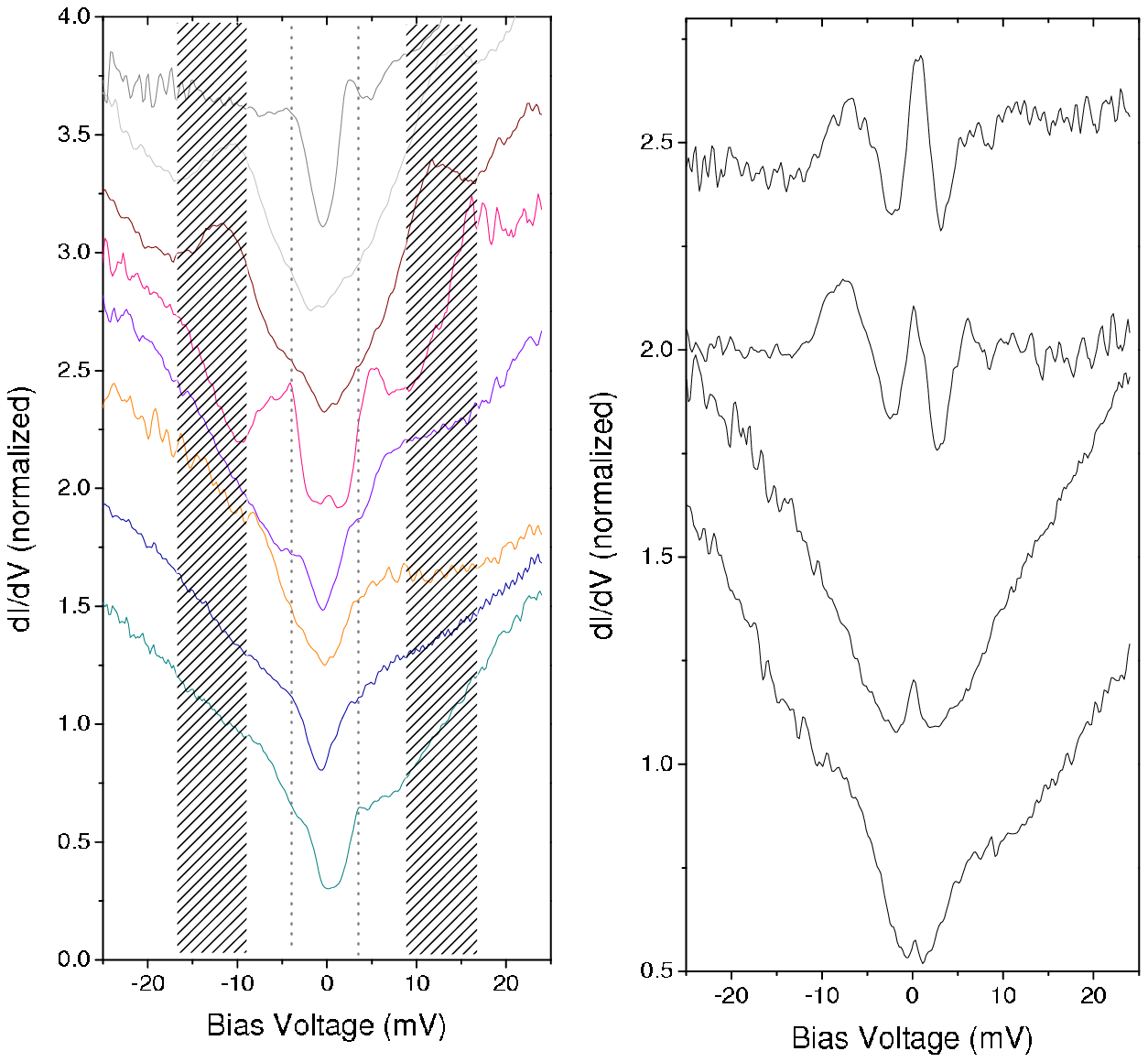}
\end{center}\caption{(Color online)
Tunneling spectra of SmFeAsO$_{0.9}$F$_{0.1}$ acquired at $T=$4.2 and at tunneling resistance $R_T$=12.5M$\Omega$ . Left panel: two gap-like structures at respectively at $\sim $11mV and $\sim $4mV a) superimposed on a 'V' shaped background are clearly visible. Dashed lines and bands are put at characteristic bias values. Right panel: selected spectra exhibiting a zero bias peak.} \label{SmFeAsOF}
\end{figure}

The tunneling conductance data observed in SmFeAsO$_{0.85}$
(Fig.\ref{SmFeAsO}(left panel)) and SmFeAsO$_{0.9}F_{0.1}$
(Fig.\ref{SmFeAsOF}(left panel)) are very different from the simple
BCS-like case of FeSe$_{0.5}$Te$_{0.5}$. A large number of spectra
exhibit two 'gap-like' structures at respectively $\sim 3-4$~meV and
$\sim 10$meV. The largest gap is defined by a kink around $\sim 10
\pm 3$meV (dashed bands in Figs.\ref{SmFeAsO},\ref{SmFeAsOF} (left
panel)) while the low energy gap is directly seen in the spectra
near zero-bias, surrounded by two peaks at $\sim 4 \pm 1$meV (dashed
line in Figs. \ref{SmFeAsO},\ref{SmFeAsOF} (left panel)). Both
spectroscopic features are observed on a 'V' shaped background
varying from one sample to other. Such a background could be related
to the electronic surface states resulting from dangling bonds or to the existence of a correlated metal. A
few spectra exhibit a zero bias peak
(Figs.\ref{SmFeAsO},\ref{SmFeAsOF} (right panel)). All spectroscopic
features we observed - two gaps, the `V' shaped background and the
presence of a zero bias peak, are in complete agreement with the
findings of Millo et al. \cite{Millo}.

\section{V. Interpretation of the results}

The two gap-like structures in the spectra of SmFeAsO$_{0.85}$ and
SmFeAsO$_{0.9}$F$_{0.1}$ strongly suggest the existence of two SC
gaps in these materials. Mazin et al. \cite{Mazin} have proposed
that a magnetic coupling could be the 'glue' for superconductivity,
which would make the pnictides very different from other multi-gap
superconductors as, for instance, the double-gap SC state in
MgB$_2$. In order to distinguish between the two situations, it is
helpful to reconsider the problem of two-gap superconductivity and,
specifically, the QP interband scattering effects.

\subsection{Two-gap model of Suhl}

As shown by Suhl, Matthias, and Walker (SMW) \cite{Suhl}, the
one-band isotropic BCS model can be extended to the case of two
energy bands. In their description, a Cooper pair can absorb a
phonon and be scattered either in the same band or in the other
band. The hamiltonian is the sum of the kinetic energy $H_K=
\sum_{k\sigma}\epsilon_kc^+_{k\sigma}c_{k\sigma}+\epsilon_k
d^+_{k\sigma} d_{k\sigma}$ and of the partial hamiltonians
describing the intraband pair scattering in each band, $H_{11}$ and
$H_{22}$ and the interband pair scattering $H_{12}$:

$$H_{11}= V_{11}\sum_{kk'}c^+_{k\uparrow}
c^+_{-k\downarrow}c_{-k'\downarrow} c_{k'\uparrow}$$
$$H_{22}= V_{22}\sum_{kk'}d^+_{k\uparrow}
d^+_{-k\downarrow}d_{-k'\downarrow} d_{k'\uparrow}$$
$$H_{12}= V_{12}\sum_{kk'}c^+_{k\uparrow}
c^+_{-k\downarrow}d_{-k'\downarrow} d_{k'\uparrow}+d^+_{k\uparrow}
d^+_{-k\downarrow}c_{-k'\downarrow}c_{k'\uparrow}$$ where $c^+$, $c$, $d^+$, $d$ are the corresponding creation and annihilation operators in each band, $V_{11}$ and $V_{22}$ are the intraband coupling potential corresponding to absorption or emission of a phonon with a pair scattering in the same band, $V_{12}$ is the interband pair coupling corresponding to the scattering of a Cooper pair from one band to the other.

This model gives rise to two gaps in the excitation spectrum,
$\Delta_1$ and $\Delta_2$, which are defined by two coupled
equations:

\begin{equation}
\left\{
\begin{array}{cc}
  \Delta_1[1-V_{11} N_1 F(\Delta_1)] =  \Delta_2 V_{12} N_2F(\Delta_2)  \\
  \Delta_2[1-V_{22} N_2 F(\Delta_2)] =  \Delta_1 V_{12} N_1F(\Delta_1)
\end{array}
\right. \label{eq_gap}
\end{equation}
where $N_\textit{i}=N_\textit{i}(E_F)$, $\textit{i}=1,2$, is the
normal state DOS at the Fermi level in each band. $F(\Delta_i)$
($i=$1,2) is a function depending on the temperature $T$, $\omega_0$
is the cutoff frequency for the mechanism responsible for the
coupling (i.e. the Debye frequency in the case of a phonon
coupling):
\begin{equation}
F(\Delta)=\int_0^{\hbar
\omega_0}\textrm{Re}\left\{\textrm{d}\epsilon
\tanh\left[\frac{\sqrt{\epsilon^2+\Delta^2}}{2k_BT}\right]/\sqrt{\epsilon^2+\Delta^2}\right\}
\end{equation}
If the interband pair coupling parameter term $V_{12}$ is neglected,
one obtains a superposition of two SC condensates with two critical
temperatures and a DOS which is the sum of two BCS-type DOS.
Otherwise ($V_{12}\neq 0$), one gets two gaps which close at the
same critical temperature depending on different parameters of the
model. This is the situation suggested for MgB$_2$ \cite{Liu_MgB2}.
However, to account for the observed SC behavior of this material
\cite{Schmidt}, the Cooper pair scattering is not enough, and the QP
scattering effects must be additionally considered \cite{Schopohl}.

\subsection{Two gap superconductor with sign reversal}

Mazin et al. \cite{Mazin} considered a more odd situation where the
pairing originates from a magnetic coupling, and the interband pair
scattering parameter $V_{12}$ becomes negative. Consequently, the
two components of the SC OP, characterized by two gaps in the
excitation spectrum, have opposite signs. Mazin et al. proposed that
such a situation could exist in the pnictides. When the intraband
coupling can be neglected compared to the interband coupling, i.e.
when $V_{11},V_{22}\ll |V_{12}|$, then the SMW Eqs.\ref{eq_gap} for
the two gaps simplify. At zero temperature, and assuming that the
cutoff frequency of the coupling mechanism is much larger than the
gap amplitude $\omega_0 \gg \Delta_{1,2}$, one obtains:
\begin{equation}
\left\{
\begin{array}{cc}
\begin{array}{cc}
  \Delta_1 = \Delta_2 V_{12} N_2 ln \left(\frac{2\hbar\omega}{\Delta_1}\right)  \\
  \Delta_2 = \Delta_1 V_{12} N_1 ln\left(\frac{2\hbar\omega}{\Delta_2}\right)
\end{array}
\end{array}
\right. \label{eq_gap_v12nulle}
\end{equation}
Therefore, it appears that in such a situation the gap in the first
band is determined mainly by the DOS in the other band. The gap
ratio at zero temperature is approximately given by $\left(
\Delta_1/\Delta_2 \right)^2\sim N_2/N_1$.

\subsection{Effect of quasiparticle scattering on the DOS}

Schopohl and Scharnberg \cite{Schopohl} considered, in addition to
SMW approximation, a term allowing the QPs to be scattered from one
band to the other. This Schopohl-Scharnberg model (SSM) of a two-gap
SC is formally equivalent to the Mc Millan model \cite{McMillan}
describing the normal metal - superconductor proximity effect in
real space. When taking into account the interband QP scattering,
not accounted for in SMW model (Eq.\,\ref{eq_gap}), the energy gaps $\Delta_i(E)$
(where $i=$1, 2) become energy dependent and are given by two
coupled equations:
\begin{eqnarray}
\Delta_1(E)&=&\frac{\Delta_1^{0}+\Gamma_{12}\Delta_2(E)/\sqrt{\Delta_2^2(E)-
(E-i\Gamma_{21})^2}}{1+\Gamma_{12}/\sqrt(\Delta_2^2(E)-(E-i\Gamma_{21})^2)} \\
\nonumber\Delta_2(E)&=&\frac{\Delta_2^{0}+\Gamma_{21}\Delta_1(E)/\sqrt{\Delta_1^2(E)-
(E-i\Gamma_{12})^2}}{1+\Gamma_{21}/\sqrt(\Delta_1^2(E)-(E-i\Gamma_{12})^2)}
\label{eqMcMillan}
\end{eqnarray}
where $\Gamma_{12}=\hbar/\tau_{12}$ and
$\Gamma_{21}=\hbar/\tau_{21}$ are the scattering coefficients and
$\tau_{ij}$ represent the QP lifetimes in each band. $\Delta_1^{0}$
and $\Delta_2^{0}$ are the gaps obtained from the self-consistency
equations:
\begin{equation}
\begin{array}{cc}
\Delta_i^{0}=\lambda_{ii}\int_0^{\hbar \omega_i}\textrm{d}E
\tanh\left[\frac{E}{2k_BT}\right]\textrm{Re}\left[\frac{\Delta_i(E)}{\sqrt{E^2-\Delta_i^2(E)}}\right]\\
+\lambda_{ij}\sqrt{\frac{N_j}{N_i}}\int_0^{\hbar\omega_{ij}}\textrm{d}E
\tanh\left[\frac{E}{2k_BT}\right]\textrm{Re}\left[\frac{\Delta_i(E)}{\sqrt{E^2-\Delta_i^2(E)}}\right]
\end{array}
\label{SelfConsistency}
\end{equation}
where i=(1,2) and $\lambda_{ii}=V_{ii}N_i$ are the intra-band
electron-phonon coupling constants, while
$\lambda_{ij}=V_{ij}\sqrt{N_iN_j}$ is the interband electron-phonon
coupling constant of the SMW model. The partial DOSs $N_S^{i}(E)$
corresponding to the two different bands $i=1,2$ are obtained by
inserting the energy dependent gaps $\Delta_1(E)$ or $\Delta_2(E)$
in the standard BCS expression for the DOS \cite{Schrieffer}:

\begin{equation}
N_S^{i}(E)=\textrm{Re}\left[\frac{|E|}{\sqrt{E^{2}-{\Delta_i(E)}^{2}}}\right]
\label{ns bcs}
\end{equation}

\begin{figure}[h]
\begin{center}
\includegraphics[width=8cm]{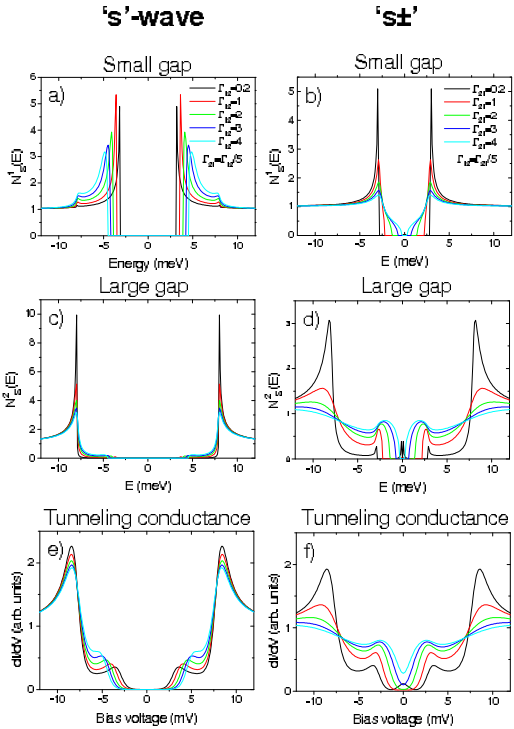}
\end{center}\caption{(Color online)
Left panel: Partial DOS, $N_S^1(E)$ (a) and $N_S^2(E)$ (c)
calculated within 's'-wave SSM model for various QP interband
scattering rates $\Gamma_{12}=$0.2, 1, 2, 3, 4meV;
$\Gamma_{21}/\Gamma_{12}=$5. The starting gap parameters are
$\Delta_1^0=$3meV, $\Delta_2^0=$8meV. (e) - corresponding tunneling
conductance calculated for $T=$4.2K using the tunneling weights
$T_1=$0.8, $T_2=$0.2. \\ Right panel: Partial DOS, $N_S^1(E)$ (b)
and $N_S^2(E)$ (d) calculated within '\textbf{s$\pm$}' SSM model for
various QP interband scattering rates. All parameters are kept the
same as in 's'-wave case (left panel) except $\Delta_1^0$ which is
of opposite sign, $\Delta_1^0=$-3meV. (f) - corresponding tunneling
conductance calculated for $T=$4.2K using the tunneling weights
$T_1=$0.8, $T_2=$0.2.} \label{SpSp_SpSm}
\end{figure}

Following Mazin et al. \cite{Mazin}, we considered that the
energy gaps in each band have opposite signs ('\textbf{s$\pm$}'
model). We further developed SSM approach to this '\textbf{s$\pm$}'
approximation.  We have then calculated the partial DOS in each
band in '\textbf{s$\pm$}' case, for different values of the
interband QP scattering parameters $\Gamma_{ij}$. We finally
compared the effects of QP scattering on SC DOS within 's'-wave
two-band SSM \cite{Noat DIPS} and '\textbf{s$\pm$}' SSM model. The
results of our calculations, done for $\Delta_1^0=\pm$3meV (small
initial gap) and $\Delta_2^0=$8meV (large initial gap), are
presented in Fig.\ref{SpSp_SpSm}, for different amplitudes of the QP
interband scattering $\Gamma_{ij}$. In the 's'-wave two-gap SSM case
(Fig.\ref{SpSp_SpSm}(left panel)), the partial DOS $N_S^{1,2}(E)$
(corresponding, respectively, to small and large gap bands 1 and 2)
are relatively weakly affected by increasing QP interband
scattering. However, in the '\textbf{s$\pm$}' SSM case
(Fig.\ref{SpSp_SpSm}(right panel)) the interband QP scattering has a
dramatic effect on the partial DOS. Both the small gap and the
large gap partial DOS ((Figs.\ref{SpSp_SpSm}(c) and (d),
respectively) show rapid filling with the states inside initial
gaps. The effect is particularly strong for the large gap DOS, since
for a high enough value of the interband QP scattering parameter
$\Gamma_{21}$, the QP peaks at the large gap energy almost disappear
in the DOS (Fig.\ref{SpSp_SpSm}(d)). In such a case, the large gap
can be hardly distinguished in the tunneling conductance spectra
(Fig.\ref{SpSp_SpSm}(f)).

In order to compare the calculated DOS to the TS data, one has to
take into account the \textit{\textbf{k}}-selectivity in the
tunneling process \cite{Mallet,Giubileo}. The simplest way is to
consider that some part $T_1$ of tunneling electrons go to the band
$i=1$ where the gap $\Delta_1(E)$ (small gap) exists whereas the
rest of electrons $T_2=1-T_1$ tunnel to the band $i=2$ characterized
by the gap $\Delta_2(E)$ (large gap). Such an approximation gives
the tunneling DOS as a weighted sum of two partial SSM DOS
(Eq.\ref{ns bcs}):
\begin{equation}
N_S^{eff}(E)=T_1 N_S^1(E)+T_2N_S^2(E),
\end{equation}
with $T_1 +T_2=1$. The tunneling conductance at finite temperature
is then obtained by replacing the term $N_S(E+eV)$ by
$N_S^{eff}(E+eV)$ in the integral $dI(V)/dV$ (Eq.\ref{eq_didv}). The
resulting curves, calculated for $T_1=0.8$, $T_2=0.2$, are presented
in Fig.\ref{SpSp_SpSm}(e) and Fig.\ref{SpSp_SpSm}(f), respectively
for 's'-wave and '\textbf{s$\pm$}' situations. We point out on a
striking difference between two results, specifically at high enough
QP scattering rates ($\Gamma_{12}\geq1$), where the large gap peaks
(at $\sim \pm8$~mV) are suppressed and only some kinks remain.
Remarkably, the QP peaks due to the small gap are very robust: They
remain clearly visible at any studied value of $\Gamma_{ij}$.

\section{VI. Discussion and perspectives}

It is tempting to compare the effects of disorder in pnictides and
chalcogenides to high-$T_c$ cuprates. In cuprates, it has been shown
that two main types of disorder can be distinguished, having
drastically different effects: in plane disorder in the CuO$_{2}$
layer and out of plane disorder (see for instance ref. \cite{Alloul}
and ref. therein). Weak disorder consists in substitution of atoms
belonging to the charge reservoir planes (\emph{out of plane
disorder}), whereas strong disorder corresponds to oxygen vacancies
or copper substitution by zinc or nickel (\emph{in plane disorder}).
The former is known to induce spatial inhomogeneities in the
superconducting state and a pseudogap in the excitation spectrum
\cite{Cren_2001}, whereas the latter is far more detrimental to
superconductivity. As an example, a few percent of Zn decreases the
critical temperature even though it is non magnetic \cite{Maeda} and
causes gapless SC \cite{Zn_cuprates}.

At first glance, the concepts developed for the cuprates can be
extented to the case of iron-based superconductors. Indeed, we can
define in the same spirit \emph{in plane} and \emph{out of plane}
disorder, with respect to the FeAs plane. However, making direct
comparison between pnictides (LnFeAsO) and chalcogenide (FeSe) is
not straightforward since, contrary to iron pnictides and cuprates,
a charge reservoir plane does not exist, strictly speaking, in the
iron chalcogenides family.

Based on the fact that the experimental spectra in SmFeAsO$_{0.85}$
and SmFeAsO$_{0.9}$F$_{0.1}$ are quite similar, we first suggest
that the SC properties are relatively insensitive to the nature of
out of plane dopants. The situation is drastically different for in
plane doping. The effect of in plane disorder is complex since, as
shown in the literature, doping by substitution depends on the
nature of the substituting atoms, isovalent or non-isovalent: The
doping can be induced in the LnFeAsO family by non-isolectronic
substitutions of iron by cobalt and nickel in the FeAs plane. In
this case, the superconducting region is dome-like with optimal
$T_{c}$ of 13~K for a doping level x$_{Co}=0.075$ in
LaFe$_{1-x}$Co$_{x}$AsO \cite{Wang_Co}\cite{Sefat_Co}. Hence, the
superconductivity appears robust with respect to strong \emph{in
plane} disorder. However, we point out that the order of magnitude
of the $T_{c}$ remains in the range of what is expected for
conventional superconductivity. On the other hand, isovalent doping
such as subtitution of As by Phosporous in LaFeAs$_{1-x}$P$_x$O
leads to superconductivity with a reduced temperature ($T_c^{mid}
\sim10.8$~K for the optimally doped $x=$0.25\textendash0.3)
\cite{Wang_EPL}. Those different results suggest that one has to
distinguish between doping effects (for non-isovalent substitution)
and disorder effects (for isovalent substitution).

Our simulations show that the effect of QP scattering in
'\textbf{s$\pm$}' case (i.e. in the case of a magnetic coupling) is
very different from the 's'-wave situation. Accordingly, the
disorder in FeAs (FeSe) plane should lead to destructive
interferences in the '\textbf{s$\pm$}' case, and only an remaining
's'-wave phononic superconductivity would survives. This point
should be further confirmed experimentally, for example, by a
decrease of the magnon frequency when the active planes are
affected. If observed, this would confirm our guess: Out of plane
defects in oxypnictides SmFeAsO$_{0.85}$ and
SmFeAsO$_{0.9}$F$_{0.1}$ do not affect strongly the SC OP and
preserve both the magnetic coupling and the two-gap structure.

In the case of Fe(Se,Te), the substitution of Se by Te should give
rise to a strong disorder, i.e. more precisely to a strong interband quasiparticle scattering that would
destroy the '\textbf{s$\pm$}' SC OP, resulting in a single
'BCS-type' SC gap (this occurs if $V_{11}$ and $V_{22}$ are non-zero).
From our full self-consistent calculations \cite{Cren Calculation}
we infer that even a very small amount of Te would probably be
enough to suppress the '\textbf{s$\pm$}' SC OP. This scenario is not
in agreement with Hanaguri et al. \cite{New Science} who claim to
observe '\textbf{s$\pm$}' superconductivity in Fe(Se,Te). The fact
that neutron scattering experiments show a resonance at 6$\sim$meV in
the magnetic scattering at the antiferromagnetic wave vector Q=(0.5,
0.5), which intensity increases abruptly when the sample is cooled
below $T_c$ \cite{MagnRes_FeSe} is also in favor of a
'\textbf{s$\pm$}' scenario \cite{footnote magnon}. The only possibility for a sign-reversal '\textbf{s$\pm$}' order parameter in FeSe$_{0.5}$Te$_{0.5}$ would be that Te substitution
does not induce strong interband quasiparticle
scattering. Nevertheless, we see a neat
difference between SmFeAsO$_{0.85}$ and FeSe$_{0.5}$Te$_{0.5}$. In SmFeAsO$_{0.85}$ we observed a strongly damped double
gap together with frequent zero-bias peaks (Figs.\ref{SmFeAsO} and
\ref{SmFeAsOF}); both are well explained in a '\textbf{s$\pm$}'
scenario. On the other hand, in the case of FeSe$_{0.5}$Te$_{0.5}$, a simple
's'-wave like gap is observed and no zero-bias peaks are present;
both facts plead for a simple BCS mechanism.

Apart from substituting Se by Te atoms, doping in FeSe is also possible 
with interplane Fe atoms, such as in Fe$_{1+\delta}$Se (which is
also sometimes written as Fe$_{x}$FeSe). Such a doping might have a
different effect. In this case, FeSe planes are preserved, the
doping being provided by the excess of Fe localized in between them.
This hypothesis seems to be confirmed by thermal conductivity
measurements by Dong et al. \cite{Dong} which are favorable to a
nodeless multigap SC OP in Fe$_{1+\delta}$Se. Moreover, the search
for a double gap structure in Fe$_{1+\delta}$Se by TS is of
immediate interest in order to understand the nature of the
superconducting coupling in this material. The impressive increase
of $T_c$ with pressure observed in Fe$_{1+\delta}$Se which can reach
37K
\cite{Mizuguchi,Medvedev_FeSe37K,Margadonna_FeSe37K,Sidorov,Miyoshi,Tissen,Garbarino},
is most probably related to the evolution of the strength of the SC
pairing mechanism (presumably of magnetic origin) with pressure. On
the other hand, the FeSe$_{1-x}$Te$_{x}$ saturates to $\sim$ 20K in
thin films with epitaxial pressure \cite{Bellingeri} and in the bulk
\cite{Sow}, a value compatible with a conventional phononic
superconductivity. The \emph{isolectronic} substitution of iron by
ruthenium and non isoelectronic substitution of iron by cobalt
atoms, which both induces a strong \emph{in plane} disorder, lead to
a dramatic decrease of T$_c$ in
NdFe$_{1-y}$M$_y$AsO$_{0.89}$F$_{0.11}$ (M=Co, Ru) which drops from
$\sim $48~K for x=0 to 0~K (non-superconducting phase) above x=0.13
\cite{Lee}. Such results strongly support our idea about the
fragility of the '\textbf{s$\pm$}' SC OP with respect to the
disorder strength.

In view of our findings, one can also expect that substitution of
iron by nickel or cobalt in LnFe$_2$As$_2$ family will lead to a
single gap structure.

\section{VII. Conclusion}

In conclusion, we have synthesized iron-based superconductors
FeSe$_{0.5}$Te$_{0.5}$ ($T_c=14$\,K), SmFeAsO$_{0.85}$ ($T_c=45$\,K)
and SmFeAsO$_{0.9}$F$_{0.1}$ ($T_c=52$\,K) and performed tunneling
spectroscopy of them at 4.2~K and in ultra-high vacuum. In
FeSe$_{0.5}$Te$_{0.5}$ crystals we observed a single 'BCS-like' gap
while in SmFeAsO$_{0.85}$, SmFeAsO$_{0.9}$F$_{0.1}$ the tunneling
spectra revealed two energy scales which we associate with a
multigap superconductivity in these materials. The tunneling
conductance spectra can be qualitatively understood within a
'\textbf{s$\pm$}' two-gap model considering the sign reversal in the
OP, due to a magnetic pairing. We showed that the '\textbf{s$\pm$}'
SC DOS is dramatically affected by QP interband scattering, in
contrast to the case of a 's'-wave two-gap superconductivity, which
was found much more robust. One of the possible explanations of the
observed single gap DOS in FeSe$_{0.5}$Te$_{0.5}$ could be in the
suppression of the '\textbf{s$\pm$}' superconductivity there by
strong disorder resulting from the Se-Te substitution.

The work was supported by French ANR grant 'GAPSUPRA'. The authors
wish to thank A. Marchenko (Institute of Physics of Kiev) for
interesting discussions as well as his precious help with the gold
samples; V. D. thanks Yves Moelo (IMN Nantes) for for interesting
discussions.


\begin{thebibliography}{00}


\bibitem{Bednorz} Bednorz J. G. and Muller K. A., Z. Phys. B. {\bf 64}, 189 (1986).

\bibitem{Kamihara} Y. Kamihara, T. Watanabe, M. Hirano, H. Hosono, J. Am. Chem. Soc. {\bf 130}, 3296 (2008).

\bibitem{Takahashi} H. Takahashi, K. Igawa, K. Arii, Y. Kamihara, M. Hirano and H. Hosono, Nature {\bf 453} 376 (2008).

\bibitem{Sm-1111}Z.-A. Ren, W. Lu, J. Yang, W. Yi, X.-L. Shen, Z.-C. Li, G.-C. Che, X.-L. Dong, L.-L. Sun, F. Zhou, Z.-X. Zhao, Chin. Phys. Lett. \textbf{25}, 6 2215
(2008).


\bibitem{review_cuprates} Patrick A. Lee, Naoto Nagaosa and Xiao-Gang Wen, Review of Modern Physics, {\bf 78}, 17 (2006).


\bibitem{Sefat_Co_PRL} Athena S. Sefat, Rongying Jin, Michael A. McGuire, Brian C. Sales, David J. Singh, and David Mandrus, Phys. Rev. Lett. {\bf  101}, 117004 (2008).

\bibitem{Torikachvili} Milton S. Torikachvili, Sergey L. Bud'ko, Ni Ni, and Paul C. Canfield, Phys. Rev. Lett. {\bf 101}, 057006 (2008)

\bibitem{Okada} H.Okada, K. Igawa, H.Takahashi, Y. Kamihara, M. Hirano, H. Hosono, K. Matsubayashi, and Y. Uwatoko, Journal of the Physical Society of Japan {\bf 77}, 113712 (2008).

\bibitem{Alireza} Patricia L. Alireza , Y. T. Chris Ko , Jack Gillett , Chiara M. Petrone , Jacqueline M. Cole , Gilbert G. Lonzarich and Suchitra E. Sebastian, J. Phys.: Condens. Matter {\bf 21} 012208 (2009).

\bibitem{Matsubayashi} K. Matsubayashi et al., N. Katayama, K. Ohgushi, A. Yamada, K. Munakata, T. Matsumoto, Y. Uwatoko, J. Phys. Soc. Jpn. {\bf 78}, 073706 (2009).

\bibitem{Igawa}  K.  Igawa, H.  Okada, H. Takahashi, S. Matsuishi, Y. Kamihara, M. Hirano, H. Hosono, K. Matsubayashi, and Y. Uwatoko, J. Phys. Soc. Jpn. {\bf 78} 025001 (2009).

\bibitem{MID} $T_c^{mid}$ is defined at the half of the resistive superconducting
transition.

\bibitem{Schopohl} N. Schopohl and K. Scharnberg, Solid State Commun. {\bf 22}
371 (1977).

\bibitem{Mazin} I. I. Mazin, D. J. Singh, M. D. Johannes, and M. H. Du, Phys. Rev. Lett. {\bf 101}, 057003 (2008).


\bibitem{Drew} A. J. Drew, F. L. Pratt, T. Lancaster, S. J. Blundell, P. J. Baker, R. H. Liu, G. Wu, X. H. Chen, I. Watanabe, V. K. Malik, A. Dubroka, K. W. Kim, M. R\"{o}ssle, and C. Bernhard, Phys. Rev. Lett.  {\bf 101}, 097010 (2008); A. J. Drew, Nature Mater. {\bf 8}, 310 (2009).

\bibitem{MagnRes_FeSe}Yiming Qiu, Wei Bao, Y. Zhao, Collin Broholm, V. Stanev, Z.
Tesanovic, Y. C. Gasparovic, S. Chang, Jin Hu, Bin Qian, Minghu
Fang, and Zhiqiang Mao, Phys. Rev. Lett. {\bf 103}, 067008 (2009);
J. Wen et al., Phys. Rev. B 81, 100513(R) (2010).

\bibitem{Christianson} A. D. Christianson et al., Nature {\bf 456}, 930 (2008).


\bibitem{Hsu} F.C. Hsu, J.Y. Luo, K.W. Yeh, T.K. Chen, T.W. Huang, P.M. Wu, Y.C. Lee, Y.L. Huang, Y.Y. Chu, D.C. Yan, M.K. Wu, Proc. Natl. Acad. Sci. USA {\bf{105}}, 14262, (2008).

\bibitem{Medvedev_FeSe37K} S. Medvedev, T. M. McQueen, I. A. Troyan, T. Palasyuk, M. I. Eremets, R. J. Cava, S. Naghavi, F. Casper, V. Ksenofontov, G. Wortmann, C. Felser, Nature mat. {\bf{8}}, 630 (2009).

\bibitem{Margadonna_FeSe37K} S. Margadonna, Y. Takabayashi, Y. Ohishi, Y. Mizuguchi, Y. Takano, T. Kagayama, T. Nakagawa, M. Takata, and K. Prassides, Phys. Rev. B {\bf 80}, 064506 (2009).


\bibitem{Sidorov} V. A. Sidorov, A. V. Tsvyashchenko and R. A. Sadykov, J. Phys.: Condens. Matter {\bf 21}, 415701 (2009).

\bibitem{Garbarino} G. Garbarino, A. Sow, P. Lejay, A. Sulpice, P. Toulemonde, M. Mezouar and
M. N\`u\~{n}ez-Regueiro, Europhys. Lett. {\bf 86} 27001 (2009).

\bibitem{Mizuguchi} Yoshikazu Mizuguchi, Fumiaki Tomioka, Shunsuke Tsuda, Takahide Yamaguchi, and Yoshihiko Takano, Appl. Phys. Lett. {\bf 93}, 152505 (2008).

\bibitem{Miyoshi} Kiyotaka Miyoshi, Yuta Takaichi, Eriko Mutou, Kenji Fujiwara, and Jun Takeuchi, J. Phys. Soc. Jpn. {\bf 78}, 093703 (2009).

\bibitem{Giaever} I. Giaever, Phys. Rev. Lett. {\bf 5},147 (1960).

\bibitem{Hess} H. F. Hess, R. B. Robinson, and
J. V. Waszczak, Phys. Rev. Lett. {\bf 64}, 2711 (1990).

\bibitem{Review_Fisher}\O ystein Fischer, M. Kugler, I. Maggio-Aprile, C. Berthod and C. Renner, Review of Modern Physics, {\bf 79}, 353 (2007).

\bibitem{Mallet} P. Mallet, D. Roditchev, W. Sacks, D. D\'{e}fourneau, and J. Klein, Phys. Rev. B, \textbf{54}, 13324 (1996).

\bibitem{Giubileo} F. Giubileo, D. Roditchev, W. Sacks, R. Lamy, D. X. Thanh, J. Klein, S. Miraglia, D. Fruchart and J. Marcus and Ph. Monod, Phys. Rev. Lett. {\bf 87}, 177008 (2001).

\bibitem{Bergeal} N. Bergeal, V. Dubost, Y. Noat, W. Sacks, D. Roditchev, N. Emery,
C. H\'erold, J-F. Mar\^ech\'e, P. Lagrange and G. Loupias, Phys.
Rev. Lett. {\bf 97}, 077003 (2006).

\bibitem{Noat_Ba8Si46} Y. Noat, T. Cren, P. Toulemonde, A. San Miguel, F. Debontridder, V. Dubost and D. Roditchev. Phys. Rev. B, \textbf{81}, 104522 (2010).


\bibitem{Kato_FeSeTe_2009} Takuya Kato, Yoshikazu Mizuguchi, Hiroshi Nakamura, Tadashi Machida, Hideaki Sakata and Yoshihiko Takano,
Phys. Rev. B {\bf 80}, 180507(R) (2009)


\bibitem{New Science} N T. Hanaguri, S. Niitaka, K. Kuroki,
H. Takagi, Science {\bf{328}}, 474 (2010).


\bibitem{Millo} Oded Millo, Itay Asulin, Ofer Yuli, Israel Felner, Zhi-An Ren, Xiao-Li Shen, Guang-Can Che, and Zhong-Xian Zhao
, Phys. Rev. B  {\bf 78}, 092505 (2008).

\bibitem{Cren_EPL2000} T. Cren, D. Roditchev, W. Sacks and J. Klein, Europhys. Lett. {\bf 52}, 203 (2000).

\bibitem{Zimmers}A. Zimmers, Y. Noat, T. Cren, W. Sacks, D. Roditchev, B. Liang, and R. L. Greene, Phys. Rev. B {\bf 76}, 132505 (2007).

\bibitem{Pan} M.H. Pan, X.B. He, G. Li, J.F. Wendelken, R. Jin, A.S. Sefat, M.A. McGuire, B.C. Sales, D. Mandrus, E.W. Plummer, arXiv:0808.0895 (2008).

\bibitem{Yin} Yi Yin, M. Zech, T. L. Williams, X. F. Wang, G. Wu, X. H. Chen and J. E. Hoffman, Phys. Rev. Lett. {\bf 102}, 097002 (2009).

\bibitem{Chen} T. Y. Chen, Z. Tesanovic, R. H. Liu, X. H. Chen and C. L. Chien, Nature {\bf 453} 1224 (2008).



\bibitem{Wang_PC} Yong-Lei Wang, Lei Shan, Lei Fang, Peng Cheng, Cong Ren and
Hai-Hu Wen, Supercond. Sci. Technol.  {\bf 22}, 015018 (2009).

\bibitem{Daghero} D. Daghero, M. Tortello, R. S. Gonnelli, V. A. Stepanov, N. D. Zhigadlo and J. Karpinski, Phys. Rev B {\bf 80}, 060502(R) (2009).

\bibitem{Liu_MgB2} Amy Y. Liu, I. I. Mazin, and Jens Kortus, Phys. Rev. Lett. {\bf 87}, 087005 (2001).

\bibitem{Schmidt} H. Schmidt, J.F. Zasadzinski, K.E. Gray and D.G. Hinks, Physica C {\bf 385}, 221 (2003).



\bibitem{Strunz} H. Strunz, E. N. Nickel, Strunz mineralogical tables, 9 th edition, E. Schweizenbart (2001).

\bibitem{RenMiniRevue} Z.-A. Ren, Z.-X. Zhao, Adv. Mater. {\bf{21}}, 4584 (2009).

\bibitem{Sefat_Co}A. S. Sefat, A. Huq, M. A. McGuire, R. Jin, B. C. Sales,D. Mandrus, L. M. D. Cranswick, P. W. Stephens, K H. Stone, Phys. Rev. B {\bf{78}}, 104505, (2008).

\bibitem{Qi}Yanpeng Qi, Zhaoshun Gao, Lei Wang, Dongliang Wang,
Xianping Zhang and Yanwei Ma, Supercond. Sci. Technol. {\bf{21}}, 115016 (2008).

\bibitem{Wang_Co}C. Wang, Y. K. Li, Z. W. Zhu, S. Jiang, X. Lin, Y. K. Luo, S. Chi, L. J. Li, Z. Ren, M. He, H. Chen, Y. T. Wang, Q. Tao, G. H. Cao, Z. A. Xu, Phys. Rev. B {\bf{79}}, 054521, (2009).

\bibitem{McQueen}T. M. McQueen, Q. Huang, V. Ksenofontov, C. Felser, Q. Xu, H. Zandbergen, Y. S. Hor, J. Allred, A. J. Williams,
D. Qu, J. Checkelsky, N. P. Ong, R. J. Cava, Phys. Rev. B {\bf{79}}, 014522 (2009)

\bibitem{Williams} A. J. Williams, T. M. McQueen, R. J. Cava, Solid State Com. {\bf{149}}, 1507 (2009).


\bibitem{Fang} M. H. Fang, H. M. Pham, B. Qian, T. J. Liu, E. K. Vehstedt, Y. Liu, L. Spinu, and Z. Q. Mao, Phys. Rev. B {\bf{78}}, 224503 (2008), K. W. Yeh, T. W. Huang, Y. L. Huang, T. K. Chen, F. C. Hsu,,Phillip M. Wu, Y. C. Lee, Y. Y. Chu, C. L. Chen, J. Y. Luo, D. C.Yan, and M. K. Wu, Europhys. Lett. {\bf{84}}, 37002 (2008).

\bibitem{Sales} B. C. Sales, A. S. Sefat, M. A. McGuire, R. Y. Jin, D. Mandrus, Y. Mozharivskyj, Phys. Rev. B {\bf{79}}, 094521 (2009)

\bibitem{Wu} M.K. Wu, F.C. Hsu, K.W. Yeh, T.W. Huang, J.Y. Luo, M.J. Wang, H.H. Chang, T.K. Chen, S.M. Rao, B.H. Mok, C.L. Chen, Y.L. Huang, C.T. Ke, P.M. Wu, A.M. Chang, C.T. Wu, T.P. Perng, Physica C  {\bf{469}}, 340 (2009)

\bibitem{Fruchart} D. Fruchart, P. Convert, P. Wolfers, R. Madar, J. P. Senateur, and R. Fruchart, Mater. Res. Bull. \textbf{10}, 169 (1975).

\bibitem{Liu} T. J. Liu, X. Ke, B. Qian, J. Hu, D. Fobes, E. K. Vehstedt, H. Pham, J. H. Yang, M. H. Fang, L. Spinu, P. Schiffer, Y. Liu,  Z. Q. Mao, Phys. Rev. B {\bf{80}}, 174509, (2009).




\bibitem{Dynes} R.C. Dynes, V. Narayanamurti, J.P. Garno, Phys. Rev.
Lett. {\bf 41}, 1509 (1978).



\bibitem{Suhl} H. Suhl, B. T. Matthias, and L. R. Walker, Phys. Rev. Lett. {\bf 3}, 552 (1959).

\bibitem{McMillan} W. L. McMillan, Phys. Rev. {\bf 175}, 537 (1968).

\bibitem{Schrieffer} J.R. Schrieffer, Rev. Mod. Phys. {\bf 200}, (1964); J.R.Schrieffer, Theory of Superconductivity, (W.A. Benjamin, New York, 1964).

\bibitem{Noat DIPS} Y. Noat, T. Cren, F. Debontridder, and D. Roditchev, W. Sacks, P. Toulemonde and A. San Miguel,  Phys. Rev. B (2010) (in press).




\bibitem{Alloul} H. Alloul, J. Bobroff,  M. Gabay and P. J. Hirschfeld, Rev. Mod. Phys. {\bf 81}, 45-108 (2009).

\bibitem{Cren_2001} T. Cren, D. Roditchev, W. Sacks and J. Klein, Europhys. Lett., {\bf  54}, 84 (2001); T. Cren, D. Roditchev, W. Sacks, J. Klein, J.-B. Moussy, C. Deville et M. Lagu\"{e}s, Phys. Rev. Lett. {\bf  84}, 1 (2000).

\bibitem{Maeda} A. Maeda, T. Yabe, S. Takebayashi, M. Hase, and K. Uchinokura, Phys. Rev. B {\bf 41}, 4112 (1990).

\bibitem{Zn_cuprates} K. Ishida, Y. Kitaoka, N. Ogata, T. Kamino, K. Asayama, J.R. Cooper and N. Athanassopoulou, Physica C {\bf 179}, 29 (1991); D. A. Bonn, S. Kamal, Kuan Zhang, Ruixing Liang, D. J. Baar, E. Klein, and W. N. Hardy, Phys Rev. B {\bf 50}, 4051 (1994).




\bibitem{Wang_EPL} Cao Wang, Shuai Jiang, Qian Tao, Zhi Ren, Yuke Li, Linjun Li, Chunmu Feng, Jianhui Dai, Guanghan Cao and Zhu-an Xu, Europhys. Lett. {\bf 86} 47002 (2009).


\bibitem{Cren Calculation} Tristan Cren et al., to be published.

\bibitem{footnote magnon} Note that such a low resonance energy of 6~meV would require a
very strong electron-magnon coupling if magnon-exchanges where
responsible of the pairing mechanism.

\bibitem{Dong} J. K. Dong, T. Y. Guan, S. Y. Zhou, X. Qiu, L. Ding, C. Zhang, U. Patel, Z. L. Xiao, and S. Y. Li, Phys. Rev. B {\bf 80}, 024518 (2009).


\bibitem{Tissen} V. G. Tissen, E. G. Ponyatovsky, M. V. Nefedova, A. N. Titov and V. V. Fedorenko, Phys. Rev. B {\bf 80}, 092507 (2009).

\bibitem{Bellingeri} E. Bellingeri, I. Pallecchi, R. Buzio, A. Gerbi, D. Marr\`{e}, M. R. Cimberle, M. Tropeano, M. Putti, A. Palenzona, and C. Ferdeghini, Appl. Phys. Lett. {\bf 96}, 102512 (2010).

\bibitem{Sow} A. Sow, P. Toulemonde, M. N\`u\~{n}ez-Regueiro, unpublished.

\bibitem{Lee} S. C. Lee, E. Satomi, Y. Kobayashi, M. Sato, Phys. Soc. Jpn {\bf{79}}, 023702 (2010).




\end{thebibliography}
\end{document}